\documentclass[twocolumn,showpacs,preprintnumbers,amsmath,amssymb,prb]{revtex4}

\usepackage{graphicx}
\usepackage{epstopdf}

\begin{document}

\title{Dynamical Mean Field Theory of Correlated Gap Formation in Pu Monochalcogenides}

\author{M.-T. Suzuki}
\author{P.\ M. Oppeneer}
\affiliation{Department of Physics and Materials Science, Box 530, Uppsala University, S-751 21 Uppsala, Sweden}

\date{\today}
\begin{abstract}
The correlated Kondo insulator state of the plutonium monochalcogenides is investigated using the dynamical mean field theory (DMFT)
 and the local density approximation +$U$ (LDA+$U$).
The DMFT-dynamical fluctuations lead to a correlated insulator state at  elevated temperature, in sharp contrast to the static LDA+$U$ approach
 that fails to reproduce both the insulating behavior and anomalous lattice constant. The DMFT conversely predicts the experimentally observed  anomalous increase of the gap with pressure
 and explains the lattice constant very well.

\end{abstract}


\maketitle

Correlated $f$-electron materials pose a major challenge to electronic structure theories. 
In particular, plutonium-based materials have received considerable attention recently because of the puzzling complexity of their correlated $5f$ electron behavior,
for which so far no unified theoretical treatment could be established.
The complex nature of Pu materials is reflected in the variety of widely different ground states that may emerge out of the competition of electronic and magnetic interactions involving the Pu $5f$ electrons.
For instance, the non-magnetic ground state of the open $5f$ shell of $\delta$-Pu is a long standing controversial issue, \cite{shick05,shim07} and
the exceptional occurrence of superconductivity at a very high critical temperature of 18.5~K discovered in PuCoGa$_5$ and its relation to the $5f$ electronic structure is an unsolved mystery.\cite{sarrao02}
A correlated insulating state that is very similar to that of Kondo insulators has been observed for the Pu monochalcogenides. \cite{fournier90,ichas01}

Energy gaps in the excitation spectrum of correlated $f$-electron materials, such as SmB$_6$ and  Ce$_3$Bi$_4$Pt$_3$, have been intensively investigated in the last years. 
\cite{riseborough00}  These Kondo insulator materials typically exhibit heavy-fermion characteristics in conjunction with narrow gap semiconducting behavior.\cite{aeppli92} 
The quasi-particle gap often shows an unusual dependence with pressure or temperature, which indicates the existence of a narrow $f$-derived feature near the top of the valence band. 
The Pu monochalcogenides show many anomalous properties that are paramount to those of Kondo insulators. In particular, they are narrow gap semiconductors with an unusual temperature and pressure dependence of the gap, \cite{ichas01,wachter91} yet the specific heat is surprisingly high, $\gamma = 30$ mJ/mol\,K$^2$. \cite{stewart91}
Although electronic structure investigations of the Pu monochalcogenides have been performed,
\cite{oppeneer00,pourovskii05,shick07} these could not provide a consistent explanation of the anomalous properties.

Here we show that the anomalous properties of the Pu monochalcogenides can be well explained by dynamical mean field theory (LDA+DMFT) calculations. In particular, the non-magnetic ground state, the formation of a correlated gap at moderate temperatures,  the anomalous increase of the gap with pressure and the unusual equilibrium lattice parameter are precisely given by LDA+DMFT theory. Remarkably, although the DMFT is conventionally considered to be ideally suited for correlated materials with a sharp quasi-particle resonance occurring at the Fermi energy, \cite{kotliar04,kotliar06} we report here the 
formation of a Kondo insulator gap within the LDA+DMFT.
We compare furthermore the results of DMFT and LDA+$U$ calculations. Although the LDA+$U$ is considered to be the static limit of the DMFT, we find that the LDA+$U$ approach {\it fails} to describe the correlated gap formation as well as the anomalous lattice constant, emphasizing the importance of dynamical fluctuations for the quasi-particle 
gap formation.

 The physical properties of the Pu monochalcogenides (PuS, PuSe, and PuTe) are experimentally well documented through transport, susceptibility, specific heat, and photoelectron spectroscopy (PES) measurements. \cite{fournier90,stewart91,gouder00,ichas01,durakiewicz04} 
 The lattice constants of the Pu monochalcogenides are anomalous, as they are too small for both divalent (5$f^6$) and trivalent (5$f^5$) Pu ions,
 which signals that these actinide chalcogenides are not plain ionic salts.
 Magnetic order, furthermore, is absent.
 \cite{fournier90}
 The temperature dependence of the resistivity signals a complex semiconducting behavior with temperature dependent gaps.\cite{fournier90}  Under pressure the room-temperature gap increases completely contrary to conventional knowledge.\cite{ichas01} The specific heat coefficient is quite high and difficult to reconcile with the semiconducting behavior. 
 Photoemission experiments performed on PuSe and PuTe reveal a sharp $f$-related peak near the Fermi edge.\cite{gouder00,durakiewicz04}

The peculiar phenomena of the Pu monochalcogenides have been attributed to intermediate valence behavior by Wachter {\it et al.} \cite{wachter91}  In this explanation the gapping near the Fermi energy is modeled through adopting a special form of the density of states (DOS). LDA calculations predict a low DOS at the Fermi level, but fail to predict the insulator state and also predict a too small equilibrium lattice parameter (Ref.\ \onlinecite{oppeneer00} and references therein). Multiplet calculations within the LDA-Hubbard I approximation (LDA-HIA) were recently used \cite{svane06,shick07} to successfully explain the peak structure present in the experimental PES spectra.
Also, an LDA+$U$ and LDA+DMFT investigation of actinide monochalcogenides was recently performed by Pourovskii {\it et al}. \cite{pourovskii05} These calculations predict an almost completely filled $5f_{5/2}$ subshell (i.e., divalent state) for the Pu monochalcogenides and do not reproduce an energy gap. 
Consequently, there is presently no theory available to suitably explain the mechanism 
driving the correlated gap formation of these Kondo insulators. We focus here on the development of this insulating state, as well as the unusual increase of the gap under pressure.

We performed electronic structure calculations using the accurate, relativistic full-potential, linearized augmented plane wave (FLAPW) method, in which the spin-orbit interaction is implemented through a second variational step. For the LDA+$U$ calculations we used the around mean field form of the double counting term. We adopt here for the screened Coulomb and exchange parameters $U= 3- 4$ eV and $J=0.55$ eV, which are in the range of commonly accepted values for Pu compounds. 
The two-electron integrals of the Coulomb interaction of the $f$ electrons are expressed in terms of  effective Slater integrals $F_{\kappa}$ ($0 \le \kappa \le 6$),  where $F_0=U$, and,
 in a manner similar to that of Ref.\ \onlinecite{shim07}, we rescale the 
 parameters. 

The ideas of the LDA+DMFT method have been reviewed previously.\cite{georges96,kotliar04,kotliar06}
We compute the local Green function 
employing the Kohn-Sham states obtained from the relativistic LDA+$U$ calculation in a manner similar to Ref.\ \onlinecite{amadon08}.
The local Green function is  computed 
on 
Matsubara frequencies $\omega_n$, through a Brillouin zone (BZ) integration of the lattice Green function, $G_{bb'} ({\mbox{\boldmath$k$}}, i \omega_n) =$ 
$\langle b{\mbox{\boldmath$k$}} |[(i\omega_n+\mu- \varepsilon_{b\boldsymbol{k}}){\bf 1} -\Delta\Sigma(i\omega_n)]^{-1}| b' {\mbox{\boldmath$k$}} \rangle$,
where $\mu$ is the chemical potential, $b {\mbox{\boldmath$k$}} $ the lattice states, and $\Delta \Sigma (i\omega_nÊ)$ the local self-energy. The latter is corrected for the static Hartree-Fock self-energy part that is already contained in the LDA+$U$ calculation, i.e., $\Delta \Sigma (i\omega_n)= \Sigma (i\omega_n)- \Sigma_{\rm HF}$. The spin-polarized T-matrix fluctuation-exchange (FLEX) impurity solver \cite{bickers89,lichtenstein98} is used here for generating the self-energy.
We note that the FLEX impurity solver is expected to be applicable to moderately correlated materials, but it cannot sufficiently resolve the atomic multiplet structure.
This implies that with the present DMFT-FLEX formulation the PES spectrum, containing atomic multiplets,\cite{svane06,shick07} cannot be fully captured. However, DMFT-FLEX  is  precisely suitable for our objective, namely, the investigation of the correlated gap formation at elevated temperatures.

In our calculations 
we used 231 $k$-points in the irreducible BZ for the self-consistent LDA+$U$ calculation and the DMFT self-consistency loop.  
In the later we take into account the $f$ orbitals and their spin, projected from all energy bands, resulting in a $14 \times 14$ matrix for the local Green function and self-energy.
Overall self-consistency  has been achieved through iterative feeding of the density matrix  of the local Green function in the next LDA+$U$ loop, and back feeding of the new solutions in the DMFT loop.
The $k$-dependent and $k$-integrated DOS was computed on 1469 $k$-points, using a Pad{\'e} approximation to  $G ({\mbox{\boldmath$k$}},i \omega) $. 
In the FLAPW calculation we included the Pu 5$f$, 6$d$, 7$s$, and 7$p$ orbitals  as band states and the Pu 5$d$, 6$s$, and 6$p$ states as semi-core states, whereas for the chalcogenides we treated the $ns$, $np$, and $nd$ states as band states and the $(n-1)d$ 
states as semi-core states, as well as the $(n-1)p$ states for PuTe.
In the DMFT calculation, we furthermore used 8192 Matsubara frequency points and applied an analytic asymptotic form  of the Green function to compensate for the remaining Matsubara frequencies when taking the sum over the frequencies.  Although we use a large number of Matsubara frequencies, we can only compute a finite number of frequency points and therefore the calculations are restricted to moderately high temperatures in practice (300 K and above). This corresponds precisely to the temperature range in which the room temperature gap of the Pu monochalcogenides has been measured.

\begin{figure}
 \begin{center}
 \includegraphics[width=0.8\linewidth]{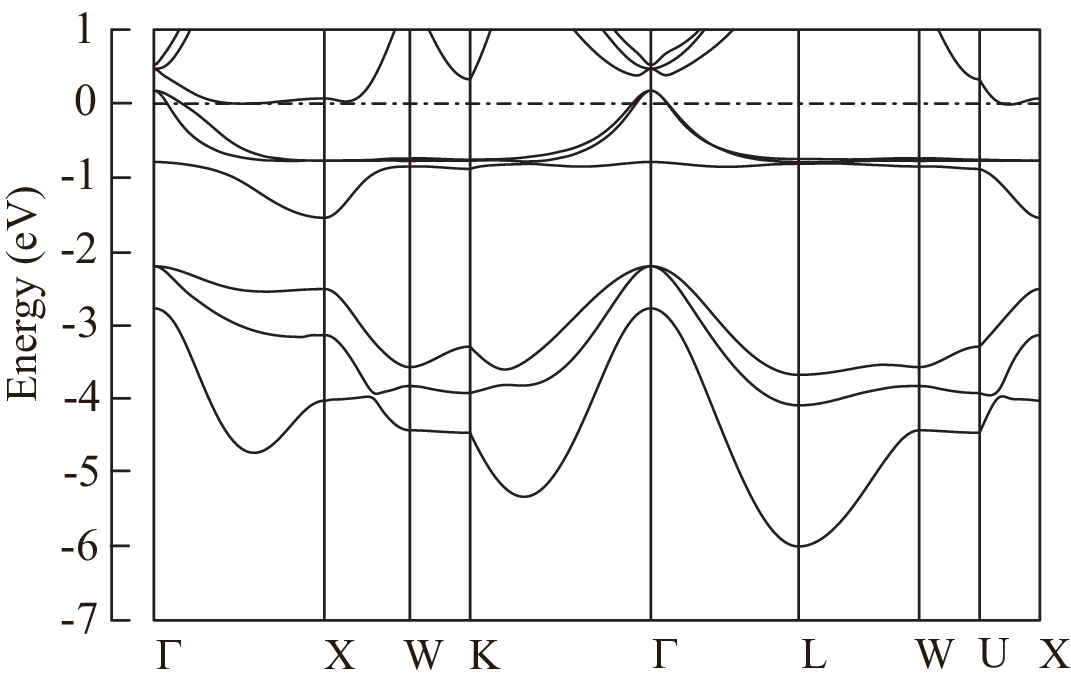}
 \includegraphics[width=0.8\linewidth]{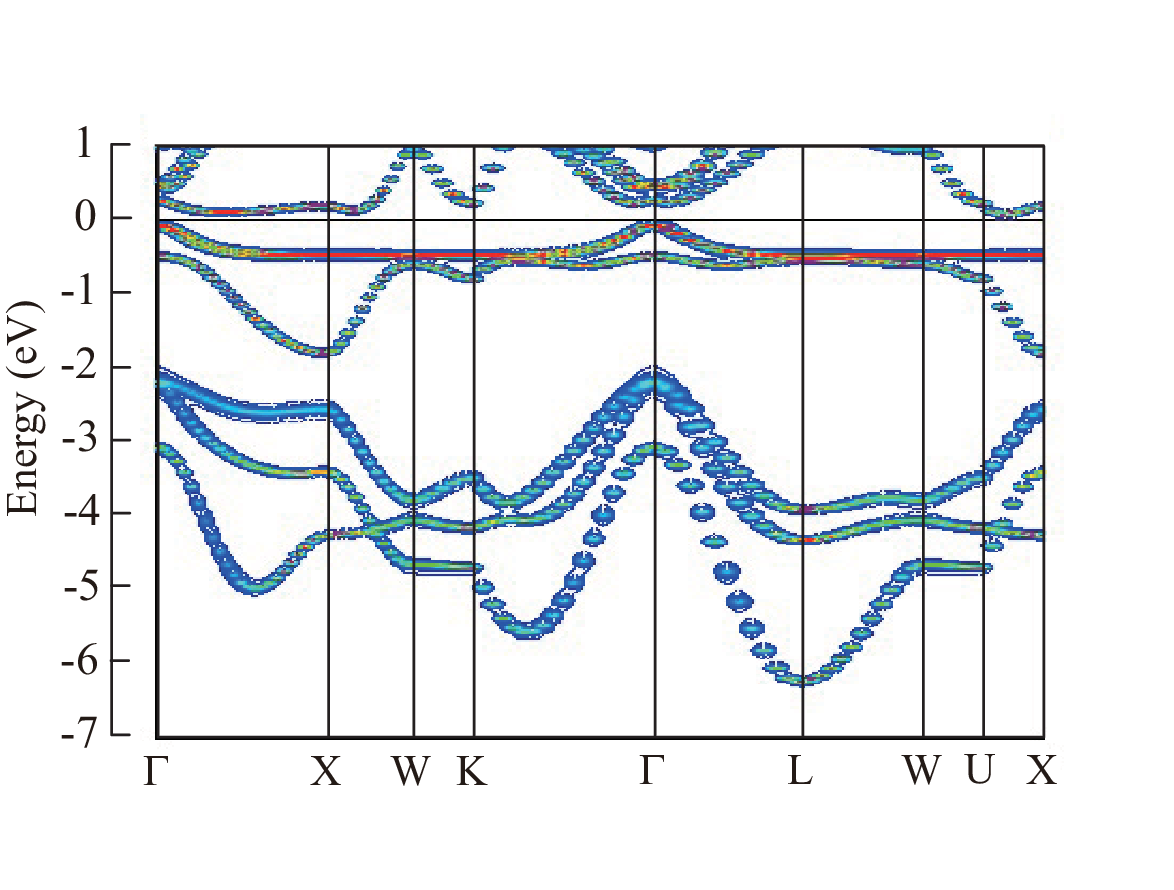}
  \caption{
  (Color online) Top: the energy bands of PuTe computed with the LDA+$U$ method at the experimental lattice constant. Bottom:  LDA+DMFT quasi-particle band structure of PuTe at $T=600$ K. The colors indicate the magnitude of the $k$-dependent spectral function,
  $-\pi^{-1}{\rm Im}\ G ({\mbox{\boldmath$k$}},E) $. }
\label{fig1}
 \end{center}
 \vspace{-0.6cm}
\end{figure}

In Fig.\ \ref{fig1} we show the  energy band structure of PuTe calculated with the LDA+$U$ method at the experimental lattice constant. As was noted before, the around mean field LDA+$U$ approach
(as well as localized limit LDA+$U$) increases the spin-orbit interaction,
favors
\textit{j-j} coupling and hence drives many Pu systems to a non-magnetic ground state.\cite{shick05,shick05b} For the Pu monochalcogenides this agrees nicely with the experimentally detected absence of magnetic ordering.\cite{fournier90}
However, as Fig.\ \ref{fig1} reveals the LDA+$U$ method predicts the Pu monochalcogenides to be semi-metallic.
Although the DOS at the Fermi level is not high, the obtained metallicity is in contradiction with experimental knowledge. This finding gives a first indication that the energy gap of the Pu monochalcogenides is special. 
For other correlated actinide compounds, such as UO$_2$ and PuO$_2$, the standard LDA approach fails to predict an insulator state, but the LDA+$U$ approach, which is designed to give the static self-energy correction to localized states, enforces the formation of a lower and an upper Hubbard band. 
However, this typical Mott-Hubbard scenario does not apply to the Pu monochalcogenides. 
In Fig.\ \ref{fig1} we also show the $k$-dependent spectral density of PuTe computed
with the LDA+DMFT.
Notably, the LDA+DMFT predicts the formation of a narrow gap in the quasi-particle spectrum.
A detailed comparison of the LDA+$U$ and LDA+DMFT bands in Fig. \ref{fig1} provides insight in the influence of the dynamical fluctuations in the self-energy that form the difference between the static LDA+$U$ and DMFT.  The dispersive  energy bands that occur at binding energies of 2 to 6 eV are dominantly due to the telluride $5p$ states with a small admixture of Pu $5f$ states; these bands are obviously unaffected. The narrow, $f$ derived band below the chemical potential 
changes more: the LDA+$U$ calculations shift this hybridized $f$ state to a lower energy, as compared to LDA calculations. Consequently, the $5f$ occupation is increased in the LDA+$U$. In the DMFT, the fluctuations within the $f$ shell are dynamically screened by fast conduction electrons, which leads to an upward shift of the narrow state. The hybridization of the $f$ states and the conduction band near the chemical potential is increased, which causes the formation of a small energy gap.
We note that the DMFT calculation, starting from the non-magnetic LDA+$U$ result, does not lead to a magnetic ground state. 
Also, 
the DMFT 
reduces the $f$ occupation with regard to that of the static LDA+$U$.

\begin{figure}
 \begin{center}
  \includegraphics[width=1.0\linewidth]{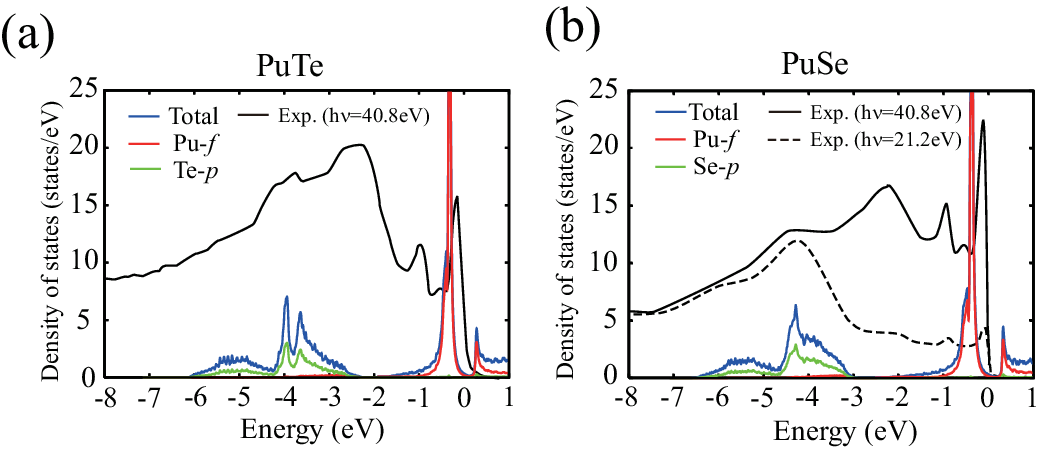}
  \caption{(Color online) The $k$-integrated DOS calculated with the LDA+DMFT method for (a) PuTe and (b) PuSe.
           The experimental photoemission spectra \cite{gouder00,durakiewicz04} are given by the solid and dashed black lines. }
\label{fig2}
 \end{center}
 \vspace{-0.6cm}
\end{figure}

In Fig.\ \ref{fig2} we show the $k$-integrated DOS obtained with the LDA+DMFT method for PuSe and PuTe.
As most of the recent experiments have been performed on PuSe and PuTe,\cite{gouder00,ichas01,durakiewicz04} we concentrate on these two chalcogenides. 
The experimental PES spectra \cite{gouder00,durakiewicz04} are also shown in Fig.\ \ref{fig2} for comparison. These experimental data contain the ``three-peak" structure at energies between 0 and $-1$ eV, followed by a broad peak at $-2$ eV and a second, smaller hump at $-4$ eV.
Considering first PuSe, we recognize that the second broad peak at  $-4$ eV is predicted to be due to the Se $p$-states. This finding is consistent with He I and He II PES \cite{gouder00} which attributes the three-peak structure and the peak at $-2$ eV to $f$-related excitations, but not the hump at $-4$ eV. Our DMFT calculations predict a narrow $f$-related peak just below the Fermi edge for PuSe and PuTe, which corresponds to a steep peak in the experimental spectra. The $f$-related peaks at $-2$, $-1$ and $-0.5$ eV have previously been attributed to multiplet structures \cite{svane06,shick07} and these are not captured by our DMFT-FLEX calculations. Interestingly, the steep peak closest to the Fermi edge has been attributed to a multiplet, too. \cite{svane06,shick07} As a salient difference, our DMFT-FLEX calculations predict this peak instead to be due to a narrow, hybridized $f$ band (see Fig.\ \ref{fig1}). Wachter \cite{wachter03} used the magnetic susceptibility to argue that the peak near the Fermi level should be due to a narrow $5f$ band, and not a multiplet peak. We propose that temperature-dependent PES measurements could be used to discriminate between these two explanations, as multiplet peaks are not expected to show a temperature dependence. 

The integrated DOS of PuSe and PuTe were also computed by Pourovskii {\it et al}., \cite{pourovskii05} however, we obtain here a DMFT solution 
that is quite different.
Pourovskii {\it et al}.\cite{pourovskii05} computed a chalcogenide DOS that is  shifted towards the Fermi edge by a considerable 2 eV as compared to our results, \cite{note} and hence,  they attribute {\it both} the experimental peaks at $-2$ and 
$-4$ eV to the chalcogenide $p$ DOS. In addition, they obtain a Pu configuration near  the closed shell divalent state, with an $f$ occupation of 5.8, both in their LDA+$U$ and non-selfconsistent  (single-shot) DMFT calculations. Our LDA+$U$ implementation conversely gives an intermediate valence $5f$ occupation of about 5.5, which, in the self-consistent LDA+DMFT loop is reduced to 5.3, as the $f$ state is depleted when it energetically comes closer to $\mu$. Thus, our DMFT calculation gives still an intermediate valence occupation, but quite close to the trivalent state, whereas the previous study \cite{pourovskii05} predicted it to be almost a $5f^6$ configuration.
Other recent DMFT studies\cite{shim07,zhu07} for $\delta$-Pu also computed Pu configurations relatively close to $5f^5$.

\begin{figure}
 \begin{center}
  \includegraphics[width=0.7\linewidth]{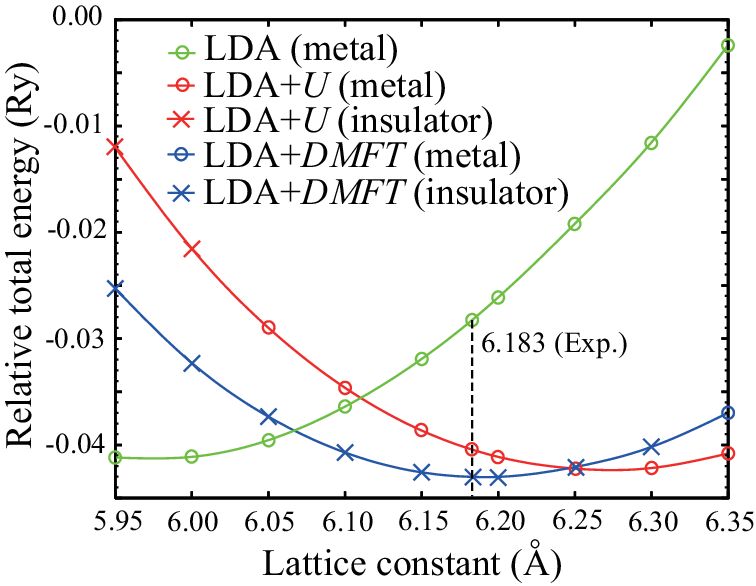}
  \caption{(Color online) The relative total energy of PuTe versus lattice constant, calculated using
   the LDA, LDA+$U$ and LDA+DMFT approaches. The experimental value is indicated by the vertical   dashed line.
Open circles and crosses indicate metallic and insulating ground states, respectively.}
\label{fig3}
 \end{center}
 \vspace{-0.6cm}
\end{figure}

The small energy gap predicted by the DMFT could be a fortuitous coincidence. We show in the following that this is not the case. First, in Fig.\ \ref{fig3} we plot the total energy of PuTe versus lattice parameter $a$, computed with the LDA, LDA+$U$, and LDA+DMFT \cite{kotliar06} approaches. The LDA predicts a metallic ground state and a much too small equilibrium lattice parameter (cf.\ Ref.\ \onlinecite{oppeneer00}). The LDA+$U$ also predicts a metallic ground state, but with a too large $a$. 
\cite{note2}
Only the LDA+DMFT predicts the insulator state {\it and} the experimental lattice constant.

\begin{figure}
 \begin{center}
  \includegraphics[width=1.0\linewidth]{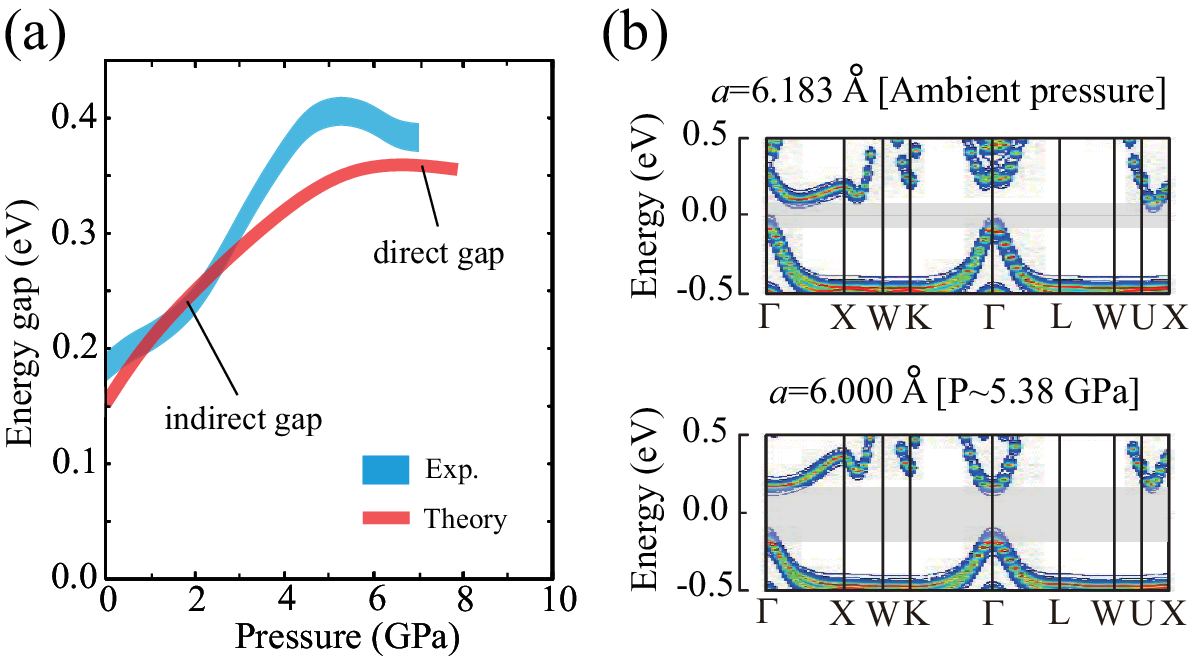}
  \caption{(Color online) (a) Pressure dependence of the energy gap of PuTe. The blue curve shows the experimental gap \cite{ichas01} and the red curve depicts the LDA+DMFT result.
 (b) The quasi-particle bands around the chemical potential for two lattice constants, 
 corresponding, respectively, to ambient pressure and $\sim$5.38 GPa.
 }
\label{fig4}
 \end{center}
  \vspace{-0.5cm}
\end{figure}

Lastly, we discuss the pressure dependence of the correlated energy gap, which is a further stringent test of the DMFT solution.
We find that pressure causes a stronger mixing of Pu $5f$ and conduction states near the Fermi level, leading to a widening of the gap. The experimental \cite{ichas01} and computed gaps of PuTe as a function of pressure are presented in Fig.\ \ref{fig4}a. 
The DMFT gap values were computed from the maximum intensity points of the highest occupied and the lowest excited state. The experiment reported that the room-temperature activation gap first increases with pressure up to 5 GPa and then slightly decreases. \cite{ichas01} Our DMFT calculations {\it completely reproduce and explain this}. The quasi-particle bands, shown in Fig.\ \ref{fig4}b, have the highest occupied state at the $\Gamma$ point. The lowest excited states are along the $\Gamma-X$ and $U-X$ directions for low pressures, but at the $\Gamma$ point for pressures above 5.38 GPa.
Thus, our calculations predict that under pressure a subtle (indirect-direct) transition in the $k$-dependent lowest excited states occurs, 
which gives rise to a turn in the slope of the gap versus pressure curve. 
The gap increase is counter-intuitively, however, it was shown that either a gap increase or decrease can occur for Kondo insulators.\cite{yuan04}

Previously there existed no explanation of the peculiar insulator state of the Pu monochalcogenides;  electronic structure approaches, as the LDA and LDA+$U$ that are based on static approximations to the self-energy, fail to reproduce the insulator state.
We have identified here that dynamical fluctuations in the DMFT are the source that drives the formation of the correlated insulator state.  Our LDA+DMFT calculations accurately explain the lattice constant as well as the anomalous increase of the room-temperature gap under pressure. 
Also, we find that the Pu $5f$ occupation, which is given to be intermediate between 5 and 6 by the LDA+$U$ approach, is reduced in our LDA+DMFT calculations, and consequently, comes closer to the trivalent configuration. Our LDA+DMFT calculations predict furthermore the presence of  a narrow Pu $5f$-derived band close to the Fermi edge. We propose that the existence of the narrow-band state could be verified through temperature-dependent PES experiments.

We thank T. Durakiewicz for helpful discussions. This work has been supported through VR, SKB,
and the Swedish National Infrastructure for Computing (SNIC). 
\vspace{-0.2cm}

\bibliographystyle{apsrev}

\begin{thebibliography}{99}
\bibitem{note}
Alternatively, one can say that the Pu $5f$ band is shifted to lower energy in Ref.\ \onlinecite{pourovskii05}.
\bibitem{note2}
At a much smaller lattice constant the LDA+$U$ also predicts a small gap, see Fig.\ \ref{fig3}. However,  this feature does not occur at a stable, equilibrium  lattice constant and, in contrast to the DMFT calculation, it happens through a hybridization of conduction bands.
\end{thebibliography}

\end{document}